\documentclass[aps,prb,twocolumn,groupedaddress]{revtex4}%
\usepackage{amsmath}
\usepackage{graphics}
\usepackage{graphicx}
\usepackage{epsfig}
\usepackage{amssymb}
\usepackage{amsmath}%
\usepackage{amsfonts}

\begin{document}
\title{Thermodynamic quantum critical behavior of the Kondo necklace model}
\author{Daniel \surname{Reyes}}
\affiliation{Centro Brasileiro de Pesquisas F\'{\i}sicas - Rua Dr.
Xavier Sigaud, 150-Urca,
 \\ 22290-180,RJ-Brazil}
\author{Mucio A. \surname{Continentino}}
\affiliation{Instituto de F\'{\i}sica, Universidade Federal Fluminense, \\
Campus da Praia Vermelha, \\
Niter\'oi, RJ, 24.210-340, Brazil}

\email{daniel@cbpf.br,mucio@if.uff.br}
\date{\today}

\begin{abstract}

We obtain the phase diagram and thermodynamic behavior of the Kondo
necklace model for arbitrary dimensions $d$ using a representation
for the localized and conduction electrons in terms of local Kondo
singlet and triplet operators. A decoupling scheme on the double
time Green's functions yields the dispersion relation for the
excitations of the system. We show that in $d\geq 3$ there is an
antiferromagnetically ordered state at finite temperatures
terminating at a quantum critical point (QCP). In 2-d, long range
magnetic order occurs only at $T=0$. The line of Neel transitions
for $d>2$ varies with the distance to the quantum critical point QCP
$|g|$ as, $T_N \propto |g|^{\psi}$ where the shift exponent
$\psi=1/(d-1)$. In the paramagnetic side of the phase diagram, the
spin gap behaves as $\Delta\approx \sqrt{|g|}$ for $d \ge 3$
consistent with the value $z=1$ found for the dynamical critical
exponent. We  also find in this region a power law temperature
dependence in the specific heat for $k_BT\gg\Delta$ and along the
non-Fermi liquid trajectory.  For $k_BT \ll\Delta$, in the so-called
Kondo spin liquid phase, the thermodynamic behavior is dominated by
an exponential temperature dependence.

\end{abstract}
\maketitle


\section{Introduction}

Quantum phase transitions from an antiferromagnetic (AF) ordered
state to a non-magnetic Fermi liquid in  heavy fermions (HF) systems
have been the subject of intense research recently\cite{Julio}. In
contrast to classical phase transitions, driven by temperature,
quantum phase transitions can be driven by magnetic field, external
pressure or doping. The physics of heavy fermion compounds is mainly
due to the competition of two main effects: the
Ruderman-Kittel-Kasuya-Yosida (RKKY) interaction between the
magnetic ions which favors long range magnetic order and the Kondo
effect which tends to screen the moments and produce a non-magnetic
ground state. These effects are contained in the Kondo Lattice Model
(KLM) Hamiltonian~\cite{Doniach} which can be derived from the more
fundamental Anderson lattice model in the case of well-developed
local spin moments \cite{Schieffer}. Although this model neglects
charge fluctuations, since the relevant QCP in heavy fermion
materials is associated with a magnetic transition, to consider only
spin fluctuations\cite{guang} turns out an excellent approximation.
The KLM has been studied by different techniques and the general
physical picture that arises is that at $T=0$, there is a quantum
phase transition from a magnetic phase at small coupling strength
$J/t$ to a non-magnetic dense Kondo phase at a critical value of
$J/t$.

Here we investigate a simplified version of this model, the so
called Kondo necklace model\cite{Doniach} (KNM), which for all
purposes can be considered as yielding results similar to the
original model. We use the bond-operator approach introduced by
Sachdev et.al.~\cite{sachdev} that was employed previously to both
KLM \cite{jure} and KNM~\cite{guang} models but always at zero
temperature. We extend this approach to finite temperatures
\cite{Da} and obtain the phase diagram of the KNM for arbitrary
dimensions. We find, that this method yields a critical Neel line in
three dimensions but not in 2-d as expected from general
arguments\cite{Mermin}. This implies that fluctuations are taken
into account to an important extent in spite of the mean-field type
of decoupling  used to deal with the bond-operator Hamiltonian. We
also calculate the thermodynamic properties along the non-Fermi
liquid (NFL) trajectory above the QCP
\cite{Mucio1,Hertz,Moriya,Millis}.

The KNM replaces the hopping term of the conduction electrons by an
$X-Y$ interaction among the conduction electron
spins~\cite{Doniach}. It is given by,

\begin{equation}
H=t\sum_{<i,j>}(\tau^{x}_{i}\tau^{x}_{j}+\tau^{y}_{i}\tau^{y}_{j})+J\sum_{i}\mathbf{S}_{i}.\mathbf{\tau}_{i},
\end{equation}
where $\tau_{i}$ and $\mathbf{S}_{i}$ are independent sets of
spin-1/2 Pauli operators, representing the conduction electron spin
and localized spin operators, respectively. The sum $\langle
i,j\rangle$ denotes summation over the nearest-neighbor sites. The
first term mimics electron propagation and in one dimension can be
mapped by the Jordan-Wigner transformation onto a band of spinless
fermions. The second term is the magnetic interaction between
conduction electrons and localized spins $\mathbf{S}_{i}$ via the
coupling $J$.

The paper is organized as follows: In Sec. II we introduce the bond
operator treatment of the KNM. In Sec. III we make and discuss the
approximations required to solve the problem. The Green's function
method is employed to attain this solution, as it can be easily
generalized to finite temperatures. We obtain a closed set of
equations for the Green's functions that can be immediately solved
without no further approximations.  The temperature dependence of
the singlet and triplet order parameters is found in Sec. IV. In
Sec. V the finite temperature phase diagram of the KNM is discussed.
In Sec. VI we study the paramagnetic phase and find the dependence
of the spin gap with the distance to QCP. In Sec. VII the behavior
of the specific heat of the KNM is presented. The last section (Sec.
VIII) offers the conclusions and discussions.

\section{ Bond operators representation}

We use the method of bond operators, which has the advantage of
making the connection to the lattice degrees of freedom most direct.
For two $S=\frac{1}{2}$ spins,  Sachdev  et. al.\cite{sachdev}
introduced four creation operators to represent the four states in
Hilbert space. This basis can be created out of the vacuum by
singlet $|s\rangle$ and triplet
$|t_{\alpha}\rangle=t_{\alpha}^{\dagger}|0\rangle$ ($\alpha=x,y,z$)
operators. In terms of these triplet and singlet operators the
localized and conduction electrons spin operators are given by,

\begin{eqnarray}
S_{i,\alpha } &=&\frac 12(s_i^{\dagger }t_{i,\alpha
}^{\phantom\dagger}+t_{i,\alpha }^{\dagger
}s_i^{\phantom\dagger}-i\epsilon _{\alpha \beta \gamma }t_{i,\beta
}^{\dagger }t_{i,\gamma }^{\phantom\dagger}),\nonumber\\
\tau _{i,\alpha } &=&\frac 12(-s_i^{\dagger }t_{i,\alpha
}^{\phantom\dagger}-t_{i,\alpha }^{\dagger
}s_i^{\phantom\dagger}-i\epsilon _{\alpha \beta \gamma }t_{i,\beta
}^{\dagger }t_{i,\gamma }^{\phantom\dagger}).
\end{eqnarray}
where $\alpha$, $\beta$ and $\gamma$ take the values $x$, $y$, $z$,
repeated indices are summed over, and $\epsilon$ is the totally
antisymmetric Levi-Civita tensor. In particular, an important
feature of the bond operator mean-field approach is that the
simplest mean-field theory to be used below already yields ground
states and excitations with the correct quantum numbers; so a strong
fluctuation analysis is not needed to capture the proper physics.
The restriction that the physical states are either singlets or
triplets leads to the constraint
$s^{\dagger}s+\sum_{\alpha}t_{\alpha}^{\dagger}t_{\alpha}=1$.
Moreover, the singlet and triplet operators at each site satisfy
bosonic commutation relations $[s,s^{\dagger }]=1, [t_{\alpha
},t_{\beta }^{\dagger }]=\delta _{\alpha ,\beta }, [s,t_{\alpha
}^{\dagger }]=0$. Substituting the operator representation of spins
defined in Eq. (2) into the original Hamiltonian and considering the
commutation relations, we obtain

\[H=H_0+H_1+H_2\]
where
\begin{eqnarray}
H_0 &=&\frac J4\sum_i(-3s_i^{\dagger
}s_i^{\phantom\dagger}+\sum_\alpha t_{i,\alpha
}^{\dagger }t_{i,\alpha }^{\phantom\dagger})\nonumber \\
&+&\sum_i\mu _i(s_i^{\dagger }s_i^{\phantom\dagger}+\sum_\alpha
t_{i,\alpha}^{\dagger }t_{i,\alpha }^{\phantom\dagger}-1),  \nonumber \\
 H_1
&=&\frac t4\sum_{\langle ij\rangle,\alpha}\left[ s_i^{\dagger
}s_j^{\dagger } t_{i,\alpha}^{\phantom\dagger}
t_{j,\alpha}^{\phantom\dagger} \right.+\left. s_i^{\dagger
}s_j^{\phantom\dagger} t_{i,\alpha}^{\phantom\dagger}
t_{j,\alpha}^{\dagger
} +h.c.\right] ,  \nonumber \\
H_2 &=&-\frac t4\sum_{\langle ij\rangle,\alpha }\left[
t_{i,z}^{\dagger }t_{j,z}^{\dagger } t_{i,\alpha}^{\phantom\dagger}
t_{j,\alpha}^{\phantom\dagger} \right.-\left. t_{i,z}^{\dagger
}t_{j,z}^{\phantom\dagger} t_{i,\alpha}^{\phantom\dagger}t_{j,\alpha}^{\dagger } +h.c.\right].\nonumber \\
\end{eqnarray}
with $\alpha=x,y$. $H_0$ represents the interaction between spins
$S$ and $\tau$ in the site $i$ and the constraint
$s^{\dagger}s+\sum_{\alpha}t_{\alpha}^{\dagger}t_{\alpha}=1$ is
implemented through the local chemical potentials $\mu_{i}$. $H_1$
and $H_2$ are terms associated with the hopping. There is still
another term in the Hamiltonian that consists of three triplets and
one singlet operator. This is not taken into account since it
vanishes in the approximation that we use below \cite{guang}.

\section{Antiferromagnetic Phase}

The Hamiltonian above,  at half filling, i.e., with one conduction
electron per site, can be simplified using a mean-field decoupling
of the quartic terms while still retaining the relevant physics. The
resulting effective Hamiltonian $H_{mf}$ with only quadratic
operators is sufficient to describe exactly the quantum phase
transition from the ordered AF state to the disordered Kondo spin
liquid at least for $d \ge 3$, as discussed below. Besides it also
yields sensible results in 2-d  where no line of finite temperature
transitions is found. Relying on the nature of the strong coupling
limit $(t/J) \rightarrow0$ we take $\langle s_i^{\dagger }\rangle
=\langle s_i\rangle =\overline{s}$, which corresponds to {\it a
condensation of the local Kondo spin singlets on each site }. Next
to describe the condensation of one local Kondo spin triplet
$t_{\mathbf{k}, x}$ on the AF reciprocal vector
$Q=(\pi/a,\pi/a,\pi/a)$, we introduce: $t_{\mathbf{k},
x}=\sqrt{N}\bar{t}\mathbf{\delta_{k,Q}}+\mathbf{\eta}_{\mathbf{k},x}$
corresponding to {\it fixing the orientation of the localized spins
along the $x$ direction}. The quantity $\bar{t}$ is the mean value
of the $x$-component spin triplet in the ground state and
$\mathbf{\eta}_{\mathbf{k},x}$ represents the fluctuations. Finally
the translation invariance of the problem implies that we may assume
the local chemical potential as a global one.

We will consider here only the terms $H_0$ and $H_1$; the $H_2$ term
has only small contributions to the results therefore it is
neglected\cite{Rice,Normand}.

After performing a Fourier transformation of the boson operators,
the mean-field effective Hamiltonian is given by,
\begin{eqnarray}
&&\hspace{0.0001cm} H_{mf}=N\left( -\frac 34J\text{
}\overline{s}^2+\mu
\overline{s}^2-\mu \right)\nonumber \\
 && +\left( \frac J4+\mu \right) \sum_{{\bf k}}t_{{\bf k},z}^{\dagger
}t_{{\bf k},z}^{\phantom\dagger} + (\frac{J}{4}+\mu-\frac{1}{2}tZ\overline{s}^2)N \bar{t}^{2}\nonumber \\
&& \hspace{0.0001cm} +\sum_{\bf k}\left[ \Lambda _{{\bf
k}}\eta_{{\bf k},x }^{\dagger }\eta_{{\bf
k},x}^{\phantom\dagger}+\Delta _{{\bf k}}\left( \eta_{{\bf k} ,x
}^{\dagger }\eta_{-{\bf k},x }^{\dagger }+\eta_{{\bf k},x
}\eta_{-{\bf k},x }\right) \right]\nonumber\\
 && \hspace{0.0001cm} +\sum_{\bf
k}\left[ \Lambda _{{\bf k}}t_{{\bf k},y }^{\dagger }t_{{\bf
k},y}^{\phantom\dagger}+\Delta _{{\bf k}}\left( t_{{\bf k} ,y
}^{\dagger }t_{-{\bf k},y }^{\dagger }+t_{{\bf k},y }t_{-{\bf k},y
}\right) \right],\nonumber\\
\end{eqnarray}
with $\Lambda _{{\bf k}}=\omega_{0}+2\Delta_{{\bf k}}$, $\lambda
({\bf k)=}\sum_{s=1}^d\cos k_s$, $\Delta _{{\bf k}}=\frac
14t\overline{s}^2\lambda ( {\bf k)}$, $N$ is the number of lattice
sites, $Z$ is the total number of the nearest neighbors on the
hyper-cubic lattice. The wave-vectors $k$ are taken in the first
Brillouin zone and the lattice spacing was assumed to be unity. This
mean-field Hamiltonian can be solved using the Green's functions to
obtain the thermal averages of the singlet and triplet correlation
functions. These are given by,

\begin{equation}
\ll t_{{\bf k},\alpha}^{\phantom\dagger};t_{{\bf
k},\alpha}^{\dag}\gg_{\omega}=\frac{1}{2\pi}\frac{\omega+\Lambda_{{\bf
k}}}{\omega^{2}-\omega_{{\bf k}}^{2}},
\end{equation}

\begin{equation}
\ll t_{{\bf k},z}^{\phantom\dagger};t_{{\bf
k},z}^{\dag}\gg_{\omega}=\frac{1}{2\pi}\frac{1}{\omega-\omega_{0}}.
\end{equation}\\
where, $\alpha=x,y$. The poles of the Green's functions determine
the excitation energies of the system as
$\omega_{0}=\left(\frac{J}{4}+\mu\right)$, which is the
dispersionless spectrum of the longitudinal spin triplet states and
$\omega_{k}=\pm \sqrt{\Lambda_{{\bf k}}^{2}-(2\Delta_{{\bf
k}})^{2}}$ that correspond to the excitation spectrum of the
transverse spin triplet states for both branches
$\omega_{x}=\omega_{y}$. From these modes and their bosonic
character an expression for the average energy at finite
temperatures can be easily obtained,
\begin{eqnarray}\label{free}
\varepsilon=\langle\mathcal{H}_{mf}\rangle&=&\varepsilon_{0}+
\frac{\omega_{0}}{2}\sum_{\mathbf{k}}\left(\coth\frac{\beta\omega_{0}}{2}-1\right)\nonumber\\
&+&
\sum_{\mathbf{k}}\omega_{\mathbf{k}}\left(\coth\frac{\beta\omega_{\mathbf{k}}}{2}-1\right),
\end{eqnarray}
where
\begin{eqnarray}\label{fund}
\varepsilon_{0}&=&N\left[-\frac{3}{4}J\overline{s}^{2}+\mu\overline{s}^{2}-\mu+\left(\frac{J}{4}+\mu-
\frac{1}{2}tZ\overline{s}^{2}\right)\overline{t}^{2} \right]\nonumber\\
&+&\sum_{\mathbf{k}}(\omega_{\mathbf{k}}-\omega_{0}),
\end{eqnarray}
is the ground state energy of the system and $\beta=1/k_{B}T$,
$\overline{s}$ and $\overline{t}$ the singlet and triplet order
parameter respectively. Since the parameter $\overline{s}$ is always
nonzero\cite{guang,jure} and $\overline{t} \ne 0$ in the
antiferromagnetic phase, we minimize the ground state energy with
respect to $\overline{t}$ to find,
$\mu=\frac{1}{2}Zt\overline{s}^{2}-J/4$ and consequently,
$\omega_{{\bf
k}}=\frac{1}{2}Zt\overline{s}^{2}\sqrt{1+2\lambda(\mathbf{k})/Z }$.
The ground state energy $\varepsilon_{0}$ corresponds to a
magnetically long-range ordered state characterized by the momentum
$q=Q$. The excitations over this ground state  are given by two
spin-wave branches associated with $t_{x}$ and $t_{y}$ that
represent rotations or transverse oscillations of the order
parameter $\bar{t}$. The third mode associated with $t_{z}$
corresponds to fluctuations in the amplitude of this order
parameter.  In the next section we consider low temperatures with
the purpose of studying the behavior of the system near the magnetic
instability.

\section{Singlet and triplet order parameters at finite temperatures}
The low temperature thermodynamic and transport properties of
heavy fermions in the vicinity of the magnetic quantum critical
point are far from being completely understood \cite{Mucio2}. In
order to study the KNM at finite temperatures we calculate in this
section the finite temperature order parameters
$\overline{s}^{2}=\overline{s}^{2}(T)$ and
$\overline{t}^{2}=\overline{t}^{2}(T)$. The free energy can
be directly obtained from the energy of the excitations given by
the poles of the Green's functions found in the previous section.
It is given by,

\begin{equation}\label{Gibbs}
G=\varepsilon_{0}-\frac{2}{\beta}\sum_{\mathbf{k}}\ln[1+n(\omega_{\mathbf{k}})]-
\frac{N}{\beta}\ln[1+n(\omega_{0})]
\end{equation}

where
\begin{equation}\label{n}
n(\omega )=\frac{1}{2}\left( \coth\frac{\beta\omega}{2}-1 \right)
\end{equation}
Minimizing the free energy of the KNM using
$(\partial\varepsilon/\partial\mu,\partial\varepsilon/\partial\overline{s})=(0,0)$,
we can easily get the following saddle-point equations,

\begin{eqnarray}\label{st}
\overline{s}^{2}&=&1+\frac{J}{Z
t}-\frac{1}{2N}\sum_{\mathbf{k}}\sqrt{1+\frac{2\lambda(\mathbf{k})}{Z}}\coth\frac{\beta\omega_{\mathbf{k}}}{2}
-\xi, \nonumber
\\
\overline{t}^{2}&=&1-\frac{J}{Z
t}-\frac{1}{2N}\sum_{\mathbf{k}}\frac{1}{\sqrt{1+\frac{2\lambda(\mathbf{k})}{Z}}}\coth\frac{\beta\omega_{\mathbf{k}}}{2}
-\xi.\nonumber\\
\end{eqnarray}
where $\xi=\frac{1}{4}(\coth\frac{\beta\omega_{0}}{2}-1)$. Generally
the equations for $\overline{s}$ and $\overline{t}$ in Eq.
(\ref{st}) should be solved and at $T=0$ the results of Ref. (4) are
recovered. For $J/t>(J/t)_c$, triplet excitations remain gapped and
at $J/t<(J/t)_c$, the ground state has both condensation of singlets
and triplets at the antiferromagnetic wave vector $Q=(\pi,\pi,\pi)$.
Then the quantum critical point at $T=0$, $(J/t)_c$ separates an
antiferromagnetic long range ordered phase from a gapped spin liquid
phase. For finite temperatures, the condensation of singlets $s$
occurs at a temperature scale which, to a first approximation,
tracks the exchange $J$. The energy scale below which the triplet
excitations condense is given by the Neel temperature ($T_N$) which
is calculated in the next section.

\section{Critical Neel line in the KNM}

In terms of the KNM, the condensation of triplets (singlets) follows
from the RKKY interaction (Kondo effect). Thus, the fact that at the
mean-field level, both $\overline{s}$ and $\overline{t}$ do not
vanish may be interpreted as the coexistence of Kondo screening and
antiferromagnetism in the ordered phase\cite{guang,jure} for all
values of the ratio $J/t<(J/t)_c$. Notice that the bond-operator
mean-field theory is appropriate in the strong coupling limit and
near the QCP. However it does not give an accurate description of
the weak limit for $(J/t) \rightarrow 0$, where the ground state is
macroscopically degenerate.

The Neel line giving the finite temperature instability of the
antiferromagnetic phase for $J/t <(J/t)_c$ is obtained as the line
in the $T$ vs $(J/t)$ plane at which $\overline{t}$
 vanishes ($\overline{t}=0$). From Eq. (\ref{st}) we can then obtain
the boundary of the AF state. From this equation we get,

\begin{equation}\label{g}
\frac{|g|}{Z}=\frac{1}{2N}\sum_{\mathbf{k}}
\frac{1}{\sqrt{1+\frac{2\lambda(\mathbf{k})}{Z}}}\left(\coth\frac{\beta\omega_{\mathbf{k}}}{2}-1\right)
+\xi,
\\
\end{equation}
where $g=|(J/t)_{c}-(J/t)|$ measures  the distance to the QCP. The
latter is given by,
$(J/t)_{c}=Z[1-\frac{1}{2N}\sum_{\mathbf{k}}\frac{1}{\sqrt{1+2\lambda(\mathbf{k})/Z}}]$
where $Z$ is the number of nearest neighbors. Expanding close to the
wave-vector $\mathbf{Q}=(\pi,\pi,\pi)$ associated with the
antiferromagnetic instability we get,

\begin{eqnarray}\label{exp}
 \lambda(\mathbf{k})=-d+\frac{k^{2}}{2} + O(k^4),
\end{eqnarray}
This  yields the spectrum of transverse spin triplet excitations as,
\begin{eqnarray}\label{spectrum}
\omega_{\mathbf{k}}=\omega_{0}\sqrt{1+
\frac{2\lambda(\mathbf{k})}{Z}}\approx \sqrt{D}k .
\end{eqnarray}
where $\sqrt{D}=\omega_{0}/\sqrt{2d}$, with $\omega_{0}$ the
z-polarized dispersionless branch of excitations and $d$ is the
Euclidean dimension. Notice that the low temperature specific heat
in the antiferromagnetic phase is dominated by the contribution of
these modes. In $3d$ this has a $T^3$ temperature dependence due to
the linear, phonon-like, dispersion relation.

Replacing Eq. (\ref{spectrum}) in Eq. (\ref{g}) and considering that
for temperatures $k_{B}T\ll\omega_{0}$, $\xi$ goes to zero faster
than the first term of Eq. (\ref{g}) we obtain,

\begin{eqnarray}\label{int}
\frac{|g|}{Z}&=&\frac{2^{d-2}S_{d}Z^{d/2}}{\pi^{d}}\left(\frac{k_{B}T}{\omega_{0}}\right)^{d-1}\nonumber\\
&\times&\int_{0}^{\frac{\pi}{2\sqrt{Z}}\frac{\omega_{0}}{k_BT}}u^{d-2}\left(\coth
u-1\right)du,\
\end{eqnarray}
where $S_{d}$ is the solid angle and $u=\beta\omega_{0}
k/2\sqrt{Z}$. For temperatures  $k_{B}T\ll\omega_{0}$ the integral
in Eq. (\ref{int}) can be calculated and we get
\begin{eqnarray}
\frac{|g|}{Z}=\frac{2^{d-2}S_{d}Z^{d/2}}{\pi^{d}\omega_{0}^{d-1}}(k_{B}T_{N})^{d-1}f(d),
\end{eqnarray}
where $f(d)=\int_{0}^{\infty}u^{d-2}\left(\coth u-1\right)du$. The
solution of this expression gives us the critical line of Neel
transitions for any dimension and temperatures $ k_{B}T \ll
\omega_{0}$ where $\omega_{0}=(\frac{J}{4}+\mu)$ is the
dispersionless spectrum of the longitudinal spin triplet states that
tracks $J$. We notice that the integral $f(d)$ diverges for $d<3$
showing that there is no critical line in two dimensions at finite
temperatures\cite{Da}, in agreement with the Mermin-Wagner
theorem\cite{Mermin}. For $d\geq3$, $f(d)$ is finite and the
equation for the critical line is given by,
\begin{equation}\label{Neel}
T_{N}\propto|g|^{\frac{1}{d-1}},
\end{equation}
which defines a shift exponent $\psi = 1/(d-1)$. In particular for
$3d$, $f(3)=\pi^2/12$ and the critical line of Neel is given by
$k_{B}T_{N}=0.13\omega_{0}|g|^{1/2}$. In summary, in this section we
have obtained analytically the expression for the Neel line close to
the QCP and we have shown  that this line does not exist for $d=2$
as expected.

\section{Paramagnetic phase}
Our interest here is to obtain the thermodynamic properties near,
but above the QCP, i.e., for finite temperatures in the spin liquid
phase for $J/t \ge (J/t)_c$. In this case we have to consider only
the condensation of singlets since $\overline{t}=0$ in the
paramagnetic phase. We find,

\begin{eqnarray}\label{para}
\varepsilon'&=&\
\varepsilon'_{0}+\frac{\omega_{0}}{2}\left(\coth\frac{\beta\omega_{0}}{2}-1\right)\nonumber\\
&+&\sum_{\mathbf{k}}\omega_{\mathbf{k}}\left(\coth\frac{\beta\omega_{\mathbf{k}}}{2}-1\right),
\end{eqnarray}
The ground state energy is now given by $\varepsilon'_{0}=N\left(
-\frac 34J\text{ }\overline{s}^2+\mu \overline{s}^2-\mu
\right)+\sum_{\mathbf{k}}\left(\omega_{\mathbf{k}}-\omega_{0}\right)$
and the free energy by,
\begin{equation}\label{Gibbspara}
G'=\varepsilon'
_{0}-\frac{2}{\beta}\sum_{\mathbf{k}}\ln[1+n(\omega_{\mathbf{k}})]-\frac{N}{\beta}\ln[1+n(\omega_{0})]
\end{equation}
The transverse spin triplet excitation spectrum can be expressed in
the general form ${\omega_{{\bf k}}}=\sqrt{\Lambda_{{\bf
k}}^{2}-(2\Delta_{{\bf k}})^{2}}$, where the excitations are triply
degenerate {\it magnons} (triplons) with a spin gap that vanishes on
approaching the QCP, as we show below.

We minimize the paramagnetic KNM free energy, deriving the following
saddle-point equations:

\begin{eqnarray}\label{auto}
2J\left(\frac{3}{4}-\frac{\mu}{J}\right)
&=&\frac{t}{N}\sum_{\mathbf{k}}\frac{\omega_{0}}{\omega_{{\bf
k}}}\lambda ({\bf k)}\coth\frac{\beta\omega_{{\bf
k}}}{2},\nonumber\\
2-\overline{s} ^2&=&\frac{1}{N}\sum_{\mathbf{k}}\frac{\Lambda _{{\bf
k}}}{\omega_{{\bf k}}}\coth\frac{\beta\omega_{{\bf k}}}{2}+2\xi.
\end{eqnarray}

At $T=0$ the self-consistent equations above are equivalent to
those found in the Kondo spin liquid ground state\cite{guang}.
Considering again the expansion close to the QCP of Eq.
(\ref{exp}), we obtain the transverse spin triplet excitation
spectrum in the paramagnetic phase as,
\begin{equation}
\label{D} {\omega_{{\bf k}}}=\sqrt{\Delta^{2}+Dk^{2}}.
\end{equation}
These quasi-particle excitations form a band whose bandwidth is a
function of $t$. The band-minimum is at $k=\pi$ and the triplet
excitation spectrum has a spin gap given by,
\begin{equation}\label{gap}
\Delta=\omega_{0}\sqrt{1-\frac{yZ}{2}},
\end{equation}
In Eq. \ref{D}, $D=y\omega_{0}^2/2$ and $y=t\overline{s}
^2/\omega_{0}$ is a dimensionless parameter. The spin gap
decreases as the hopping $t$ (hence $y$) increases and vanishes
when $y\rightarrow2/Z$.

\subsection{Dependence of the spin gap on the distance to the QCP}
We now obtain the dependence of the spin gap $\Delta$ with the
distance to QCP, given by $g=|J/t-(J/t)_{c}|$, in the quantum
disordered Kondo spin liquid state from Eq. (\ref{auto}) at $T=0$.
This was done previously for the spin gap in spin ladder systems but
always in one dimension\cite{Rice,Dagotto}.

The first step is to express Eq. (\ref{auto}) as an equation for the
parameter $y$,
\begin{equation}\label{y}
y=2\left(\frac{t}{J}\right)\left[1-\frac{S_{d}}{2^{1-d}\pi^{d}}\frac{1}{\sqrt{1+yd}}K(\eta)\right],
\end{equation}
with
\begin{equation}\label{K}
K(\eta)=\int_{0}^{\pi/2}\frac{\theta^{d-1}d\theta}{\sqrt{1-\eta^{2}\sin^{2}{\theta}}},
\end{equation}
where $\eta=\sqrt{2yd/(yd+1)}$ and $K(\eta)$ in one dimension is the
complete elliptic integral of the first kind\cite{Rice}. In Eq.
(\ref{gap}) we see that when $y= y_c = 2/Z$, the spin gap vanishes
yielding
\begin{equation}\label{QCP}
\left(\frac{J}{t}\right)_{c}=\frac{2}{y_c}\left[1-\frac{S_{d}}{2^{1-d}\pi^{d}}\frac{1}{\sqrt{1+y_c
d}}K\left(\eta_c\right)\right],
\end{equation}
where $\eta_c=\sqrt{2y_cd/(y_cd+1)}$. Since $|g|=|J/t-(J/t)_{c}|$ we
obtain,

\begin{equation}\label{x-y}
\frac{2(y_c-y)}{y_cy}=|g|+\frac{S_{d}2^{d}}{\pi^{d}}G(y),
\end{equation}
where
\begin{equation}
G(y)=\frac{1}{y\sqrt{1+yd}}K(\eta)-\frac{1}{y_c\sqrt{1+y_cd}}K(\eta_c),
\end{equation}
The above equation can be expanded in Taylor's series for $y$ near
$y_c$ yielding,
\begin{equation}\label{Taylor}
G(y)\approx \frac{\partial}{\partial y}
\frac{1}{y\sqrt{1+yd}}K(\eta)|_{y=y_c}(y-y_c)+O(y-y_c)^{2},
\end{equation}

Now considering Eq. (\ref{x-y}) we can express the spin gap energy
by,
\begin{equation}\label{Gap}
\Delta=J\left(\frac{1}{4}+\frac{\mu}{J}\right)\sqrt{\frac{yZ|g|}{2\left(Z+\zeta
(y_c,d) \right)}},
\end{equation}
where $\zeta(y_c,d)=\frac{S_d 2^{d}}{\pi^d}\frac{\partial}{\partial
y} \frac{1}{\sqrt{1+yd}}K(\eta)|_{y=y_c}$ with $\zeta(y_c,2) \approx
297.01$ and $\zeta(y_c,3)\approx 17.78$ in two and three dimensions
respectively. $\mu/J$ and $y$ are given by the self-consistent
equations (\ref{auto}) and (\ref{y}), respectively and are never
critical. This relation between the spin gap and $|g|$ shows that
when $(J/t)$ decreases  from its strong coupling limit, the triplet
gap at the wave vector $k=(\pi,\pi,\pi)$ decreases, and vanishes at
$|g|=0$. Since $\Delta \propto |g|^{1/2}$, when $|g| \rightarrow 0$,
we can immediately identify the gap exponent $\nu z=1/2$ at the
quantum critical point of the Kondo lattice.

\section{Specific Heat in the Paramagnetic phase}

The purpose of this section is to consider the paramagnetic
half-filled KNM energy given by Eq. (\ref{Gibbspara}) and discuss
its finite-temperature properties near the QCP. Already the ground
state and finite-temperature properties of the KLM have been
intensively studied for one dimension using the density matrix
renormalization group (DMRG) method\cite{Shi,Shi1,Shi2}. For two
dimensions this has been accomplished using finite-temperature
Lanczos technique\cite{Haule}, perturbation theory\cite{Gu} and
quantum Monte Carlo (QMC) simulation\cite{Assad}. Lately, also for
the KNM using QMC for two dimensions\cite{Assad1}. However, less
analytical work has been carried out for higher-dimensional KNM and
KLM. This way, motivated by the above, we are going to analyze the
specific heat $C/T \equiv \gamma$ using an analytical treatment
valid for any dimension. We obtain $\gamma$ in the paramagnetic
disordered phase, in two cases: First in the so called non-Fermi
liquid trajectory ($|g|=0, T\ne 0$) and then, below the spin gap for
$k_B T \ll \Delta\propto|g|^{1/2}$. We consider the free energy Eq.
(\ref{Gibbspara}) with $\omega_{{\bf k}}=\sqrt{\Delta^{2}+Dk^{2}}$
and take into account only the contribution of the transverse spin
triplet excitation spectrum $\omega_{{\bf k}}$, since the
contribution of longitudinal triplet excitation $\omega_{0}$ tends
to zero faster at low temperatures $k_{B}T\ll\omega_{0}\sim J $.

\subsection{Specific heat along the NFL trajectory}

Non Fermi-liquid behavior is often found near a magnetic
QCP\cite{Stewart}, indicating that the NFL state in those systems
may be linked to the magnetic instability at $T=0$. In this state
the coefficient of the linear term of the specific heat does not
saturate as expected from the Landau scenario but shows a
temperature dependence as the temperature is lowered. Exactly at the
QCP, this may occur down to the lowest temperatures \cite{livroM}.
Next, we are going to find analytically $C/T$ at the NFL trajectory.
From Eq. (\ref{Gibbspara}) and using $C/T=-\partial^2 G'/\partial
T^2$ we get,

\begin{equation}\label{heat}
C/T=\frac{k_{B}^2S_{d}}{D^{d/2}2\pi^{d}}\varsigma(d)(k_{B}T)^{d-1},
\end{equation}
 where
\begin{equation}
\varsigma(d)=\int_{x_{0}}^{\sqrt{x_{0}^{2}+
D\beta^{2}\pi^{2}}}u^3(u^2-x_{0}^2)^{\frac{d-2}{2}}(\coth^2\frac{u}{2}-1)du,\nonumber\\
\end{equation}
with $x_{0}=\Delta/k_{B}T$,
$\Delta=\omega_0\sqrt{yZ|g|/[2\left(Z+\zeta (y_c,d) \right)]}$,
$u=\sqrt{x_{0}^2+D\beta^2 k^2}$ and $g=|(J/T)-(J/T)_{c}|$ measures
the distance to the QCP. Equation (\ref{heat}) yields the
expression for the specific heat in the paramagnetic region and it
is due, as already we pointed out, to the contribution of bosons
$t_x$ and $t_y$. At $|g|=0$ and low temperatures, such that, $
J\gg k_{B}T\gg \Delta$ (NFL trajectory) we can rewrite Eq.
(\ref{heat}) as follows,
\begin{equation}\label{NFL}
C/T|_{|g|=0}=\gamma_{|g|=0}=\frac{S_{d}k_{B}^2\varsigma(|g|=0,d)}{D^{d/2}2\pi^{d}}(k_{B}T)^{d-1},
\end{equation}
where $\varsigma(|g|=0,d)=\int_{0}^{\infty}u^{d+1}(\coth^2
\frac{u}{2}-1)d u$ is $24\zeta(3)$ and $16\pi^4/15$ in two and three
dimensions, respectively and $\zeta$ is the Riemann zeta-function.
Thereby $C/T\propto T^{d-1}$ at the QCP. Notice that this is
consistent with the general scaling result $C/T \propto T^{(d-z)/z}$
with the dynamic exponent taking the value \cite{livroM} $z=1$ .
Together with the previous result for the gap exponent $\nu z=1/2$,
we find that $\nu=1/2$, a result in agreement with the mean-field or
Gaussian character of the approximations we have used to deal with
the bond-operator Hamiltonian. Furthermore, since $z=1$, in three
dimensions $d_{eff}=d+z=d_c=4$ where $d_c$ is the upper critical
dimension for the magnetic transition \cite{livroM}. Consequently,
the present approach yields the correct description of the quantum
critical point of the KNM for $d\ge 3$.

\begin{figure}[th]
\centering \includegraphics[angle=0,scale=1.30]{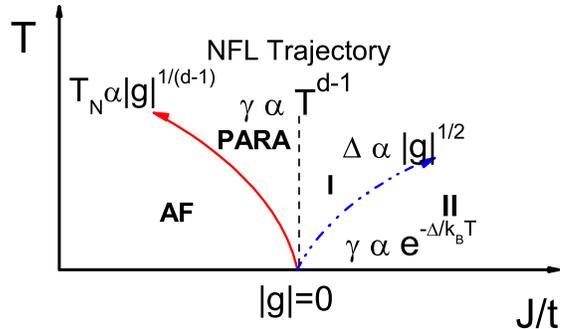}
\caption{(Color online) Schematic phase diagram of the Kondo
necklace model at finite temperatures. The AF phase is located below
the Neel line $T_N$ (solid line), which vanishes at a critical value
$(J/t)_c$ (QCP). The spin gap energy (dashed line) is shown as a
function of $g$. Below this line there is a Kondo spin liquid state.
The temperature dependence of the specific heat
\protect{$\gamma=C/T$} is given along the NFL trajectory and in the
Kondo spin liquid state for any dimension. The phase diagram shows
that at low temperatures a small value of $J/t$ supports the
development of AF order via the RKKY interaction. If the value of
$J/t$ becomes too high, the local moments are quenched by the
conduction electrons and the system has a dense Kondo ground state.}
\end{figure}

\subsection{Specific heat in the Kondo spin liquid state}

In the region II of the Fig. (1) we have to consider the
paramagnetic contribution to the specific heat given by Eq.
(\ref{para}) but now taking into account that $k_{B}T\ll \Delta$.
Then for $k_{B}T \ll \Delta\approx\sqrt{|g|}$ we obtain,

\begin{equation}
C/T=\frac{2S_{d}}{\pi^{d}D^{d/2}}
\frac{\Delta^{d+2}}{k_BT^{3}}e^{\Delta/ k_BT}\delta(d,T),
\end{equation}
where $\cosh u_{f}=1+x_{0}^{-1}$, $x_{0}=\Delta/k_{B}T$ and
$\delta(d)=\int_{0}^{u_{f}}(\sinh u)^{d-1}\cosh^{3}u du$. This can
be calculated and we obtain
$\delta(d=2,T)=\left(\frac{k_BT}{\Delta}+\frac{3}{2}(\frac{k_BT}{\Delta})^2\right)$
and
$\delta(d=3,T)=\frac{\sqrt{2}}{3}\left(2(\frac{k_BT}{\Delta})^{3/2}+\frac{17}{5}(\frac{k_BT}{\Delta})^{5/2}\right)$
in two and three dimensions respectively. In this case the specific
heat is governed by the exponential term and  we can conclude that
the dimensionality does not make much difference in thermodynamic
properties. This point was also indicated using numerical methods
for the $2d$ KLM\cite{Haule}.

Summarizing we have

\begin{equation}\label{summ}
C/T  \propto \left\{ \begin{array}{ll}
             T^{d-1}&\mbox{ NFL trajectory, $J \gg k_{B}T\gg
             \Delta$}\\
               e^{-\Delta/k_{B}T}& \mbox{,
$k_{B}T\ll
               \Delta$}\vspace{.2cm}\\
          \end{array}
        \right.,
\end{equation}
where in the latter case only the dominant term has been written.
Then, we have obtained the specific heat due to the magnetic degrees
of freedom in the paramagnetic phase of the half-filled KNM for any
dimension and temperatures $k_BT\ll J$. We have investigated it in
two cases: along the NFL trajectory and in the dense Kondo spin
liquid phase where an exponential temperature due to the spin gap
dominates over the power law associated with different
dimensionalities. Unfortunately, there are few existing results in
higher dimension considering the approaches done; almost all results
above were obtained for the strong coupling limit ($t=0$). Our work
should provide a valuable benchmark for other approaches. While we
have not attempted a detailed comparison on or fit to experiment,
this would be possible.

\section{Conclusions}

In the present work, we have constructed analytical expressions to
find the phase diagram of the Kondo necklace model for any dimension
and at low temperatures, such that, $k_BT\ll J$. Although there were
several approaches treating the Kondo lattice model with methods
similar to the one we use, they were all restricted to zero
temperature. The present finite temperature treatment allowed us to
determine the critical line for antiferromagnetic transitions, as
well as, the thermodynamic behavior along the quantum critical
trajectory and in the non-critical part of the phase diagram. We
obtained the relevant critical exponents governing the critical line
and the thermodynamic behavior due to the magnetic degrees of
freedom. For three dimensions these results turn out to be exact
since $d_{eff}=d+z=4$ is equal to the upper critical dimension for
the Kondo lattice.

We also found that the spin gap in the Kondo singlet phase vanishes
as $\sqrt{|g|}$ at the QCP consistent with $z=1$ and a mean field
correlation length $\nu=1/2$.

\acknowledgments The authors would like to thank Prof A. Troper for
helpful discussions and also the Brazilian Agencies, FAPERJ and CNPq
for financial support.

\end{document}